\documentclass[11pt]{article}
\pdfoutput=1 
\usepackage{setspace}
\usepackage{multirow}
\usepackage[table,xcdraw]{xcolor}
\usepackage{colortbl} 
\usepackage[table,xcdraw]{xcolor}
\usepackage{graphicx}
\usepackage{amsmath}
\usepackage{float}

\usepackage[T1]{fontenc}
\usepackage{amssymb}
\usepackage{eucal} 
\usepackage[sc,osf]{mathpazo}

\usepackage{xcolor,colortbl}
\usepackage{forloop}

\usepackage{hyperref}
\hypersetup{pageanchor,breaklinks,plainpages=false,colorlinks=true,citecolor=green!50!black,linkcolor=red!70!black}

\usepackage[margin=1in,headsep=0.1cm,footskip=0.0in,includeheadfoot]{geometry}  
\usepackage{graphicx}
\usepackage{amsmath}
\usepackage{listings}
\usepackage{sectsty}
\usepackage{enumerate}
\usepackage{comment}
\usepackage{wrapfig}
\usepackage{caption}
\usepackage{blindtext}
\usepackage{booktabs}

\usepackage{tikz,circuitikz}
\usetikzlibrary{patterns,calc,angles,quotes,automata,positioning,decorations.markings,arrows,intersections,circuits.logic.US,shapes.arrows,shapes.gates.logic.US,backgrounds}
\usetikzlibrary{automata,positioning,decorations.markings,arrows,intersections,calc,shapes}

\colorlet{darkgreen}{green!40!black}
\colorlet{darkblue}{blue!60!black}
\colorlet{darkred}{red!50!black}
\colorlet{safecellcolor}{yellow!5}
\colorlet{goodcellcolor}{green!10}
\colorlet{badcellcolor}{blue!10}

\tikzset{
  >=stealth,
  box state/.style={draw,rectangle,minimum size=8mm},
  prob state/.style={draw,very thick,shape=circle,darkblue,minimum size=3mm,inner sep=0mm},
  node distance=2cm,on grid,auto, initial text=,
  every loop/.style={shorten >=0pt},
  accepting/.style={double distance=1.2pt, outer sep = 0.6pt+\pgflinewidth},
  accepting dot/.style={above=-2.5pt,circle,fill,darkgreen,inner sep=2pt,radius=1pt},
  loop above/.append style={every loop/.append style={out=120, in=60, looseness=6}},
  loop below/.append style={every loop/.append style={out=300, in=240, looseness=6}},
  loop left/.append style={every loop/.append style={out=210, in=150, looseness=6}},
  loop right/.append style={every loop/.append style={out=30, in=330, looseness=6}},
  accepting arc/.style={dashed},
  marked/.style={
    dashed,
    opacity=0.3
  },
  marked on/.style={alt=#1{marked}{}},
}
\ctikzset{bipoles/not port/width=0.75,bipoles/not port/height=0.5}
\ctikzset{bipoles/not port/circle width=0.3}
\ctikzset{tripoles/american nand port/circle width=0.2}

\definecolor{lightgray}{rgb}{0.98,0.98,0.98}

\usepackage{ntheorem}

\renewtheorem*{theorem*}{Theorem}

\usepackage{mathrsfs}

\definecolor{darkgreen}{rgb}{0,0.6,0}
\definecolor{lightblue}{rgb}{0.5,0.6,1.0}
\definecolor{mauve}{rgb}{0.58,0,0.82}
\definecolor{sienna}{rgb}{0.6,0.18,0.09}
\colorlet{darkblue}{blue!60!black}
\colorlet{darkred}{red!50!black}
\colorlet{safecellcolor}{yellow!5}
\colorlet{goodcellcolor}{green!10}
\colorlet{badcellcolor}{blue!10}

\newcommand{\stsays}[1]{{\bf ST:} \textcolor{blue}{#1}}

\usepackage{xcolor}

\usepackage{color}

\definecolor{dkgreen}{rgb}{0,0.6,0}
\definecolor{gray}{rgb}{0.5,0.5,0.5}
\definecolor{mauve}{rgb}{0.58,0,0.82}
\definecolor{deepred}{rgb}{0.6,0,0}
\definecolor{deepgreen}{rgb}{0,0.5,0}

\usepackage{listings}

\DeclareFixedFont{\ttm}{T1}{txtt}{m}{n}{9}  
\DeclareFixedFont{\ttb}{T1}{txtt}{bx}{n}{9}

\usepackage[tikz]{bclogo}
\usepackage[framemethod=tikz]{mdframed}
\usepackage{lipsum}
\usepackage[many]{tcolorbox}

\usepackage[noframe]{showframe}
\usepackage{framed}

 {\endMakeFramed}
 \definecolor{shadecolor}{gray}{0.85}

\definecolor{bgblue}{RGB}{245,243,253}
\definecolor{ttblue}{RGB}{91,194,224}

\mdfdefinestyle{mystyle}{%
  rightline=true,
  innerleftmargin=10,
  innerrightmargin=10,
  outerlinewidth=3pt,
  topline=false,
  rightline=false,
  bottomline=false,
  skipabove=\topsep,
  skipbelow=false
}

\newtcolorbox{myboxi}[1][]{
  breakable,
  title=#1,
  colback=white,
  colbacktitle=white,
  coltitle=black,
  fonttitle=\bfseries,
  bottomrule=0pt,
  toprule=0pt,
  leftrule=3pt,
  rightrule=3pt,
  titlerule=0pt,
  arc=0pt,
  outer arc=0pt,
  colframe=black!50,
}

\newtcolorbox{myboxii}[1][style=mystyle]{
  breakable,
  freelance,
  colback=white,
  colbacktitle=white,
  coltitle=black,
  fonttitle=\bfseries,
  bottomrule=0pt,
  boxrule=0pt,
  colframe=white,
  after skip=0pt,
  overlay unbroken and first={
    \draw[white!75!black,line width=3pt]
    ([yshift=-9pt]frame.north west) --
    ([yshift=9pt]frame.south west);
  },
  }

\newtcolorbox{myboxiii}[1][style=mystyle]{
  breakable,
  freelance,
  colback=white,
  colbacktitle=white,
  coltitle=black,
  fonttitle=\bfseries,
  bottomrule=0pt,
  boxrule=0pt,
  colframe=white,
  after skip=0pt,
  overlay unbroken and first={
    \draw[red!75!black,line width=3pt]
    ([yshift=-9pt]frame.north west) --
    ([yshift=9pt]frame.south west);
  },
  }
\newtcolorbox{myboxv}[1][style=mystyle]{
  breakable,
  freelance,
  colback=white,
  colbacktitle=white,
  coltitle=black,
  fonttitle=\bfseries,
  bottomrule=0pt,
  boxrule=0pt,
  colframe=white,
  after skip=0pt,
  overlay unbroken and first={
    \draw[green!75!black,line width=3pt]
    ([yshift=-9pt]frame.north west) --
    ([yshift=9pt]frame.south west);
  },
  }

\usepackage[mathlines]{lineno}

\usepackage{subcaption}

\newmdenv[shadow=false, shadowsize=0pt,
linewidth=2pt, frametitlerule=false, roundcorner=10pt,frametitlebackgroundcolor=black!10]{myshadowbox}

\usepackage{fancyhdr}
\tikzstyle{loc}=[draw,circle,fill=white, minimum size=1.5em,inner sep=0em]
\tikzstyle{port}=[draw,circle,fill=white, minimum size=.35em,inner sep=0em]
\tikzstyle{probloc}=[draw,circle,fill=black!40, minimum size=.35em,inner sep=0em]
\tikzstyle{oloc}=[draw,rectangle,fill=black, minimum size=2em,inner sep=0em]
\tikzstyle{oport}=[draw,rectangle,fill=black!90, minimum size=.5em,inner sep=0em]
\tikzstyle{boxloc}=[draw, fill=yellow!10, very thick, rectangle, minimum size=2em, inner sep=0.3em]
\tikzstyle{trans}=[-latex, rounded corners]
\tikzstyle{trans2}=[-latex, rounded corners, dashed]
\tikzstyle{rate}=[fill=gray!10,rounded corners,fill opacity=0.9]
\tikzstyle{inv}=[]

\colorlet{darkgreen}{green!40!black}
\colorlet{darkblue}{blue!60!black}
\colorlet{darkred}{red!50!black}
\colorlet{safecellcolor}{yellow!5}
\colorlet{goodcellcolor}{green!10}
\colorlet{badcellcolor}{blue!10}

\usepackage{makecell}
\usepackage[symbol]{footmisc}
\usepackage{fancyhdr}

\begin{document}

\section*{Technical Challenges in Maintaining Tax Prep Software with \\ Large Language Models\footnote{\href{https://taxpolicycenter.org/event/14th-annual-irstpc-joint-research-conference-tax-administration}{14th Annual IRS/TPC Joint Research Conference on Tax Administration (IRS-TPC 2024).}}}

\noindent Sina Gogani-Khiabani, UT El Paso, USA

\noindent Varsha Dewangan, CU Boulder, USA 

\noindent Nina Olson, Center for Taxpayer Rights, USA

\noindent Ashutosh Trivedi, CU Boulder, USA

\noindent Saeid Tizpaz-Niari\footnote{\textbf{Corresponding Author}: saeid@uic.edu}, UT El Paso, USA

\subsection*{Abstract}
{\small As the US tax law evolves to adapt to ever-changing politico-economic realities, tax preparation software plays a significant role in helping taxpayers navigate these complexities. 
The dynamic nature of tax regulations poses a significant challenge to accurately and timely maintaining tax software artifacts. 
The state-of-the-art in maintaining tax prep software  is time-consuming and error-prone as it involves manual code analysis combined with an expert interpretation of tax law amendments. 
We posit that the rigor and formality of tax amendment language, as expressed in IRS publications, makes it amenable to automatic translation to executable specifications (code). 
Our research efforts focus on identifying, understanding, and tackling technical challenges in leveraging Large Language Models (LLMs), such as ChatGPT and Llama,
to faithfully extract code differentials from IRS publications and automatically integrate them with the prior version of the code to automate tax prep software maintenance. }


\section{Introduction}
\label{sec:introduction}

The growing complexity of US income tax laws has made manual tax return preparation burdensome and susceptible to errors. According to the IRS, 90 percent of tax filers submitted their taxes electronically in 2020~\cite{IRS-efile}. Additionally, the use of software for tax preparation is on the rise, with more than 72 million individuals preparing their taxes independently, without tax professionals, marking a 24 percent increase from 2019~\cite{IRS-rates}. 
As a result, the industry revenue for tax preparation services has grown to an estimated \$13.9 billion in 2023~\cite{market-size}.
Recently, the IRS introduced Direct File, an 
online software tool that provides free online tax filing in 12 states~\cite{IRS-directfile}. 

The development of socio-legal critical software is known to be challenging~\cite{EscherB20}, as it requires combined expertise in mission-critical software development practices and legal framework interpretation. 
The ever-changing nature of tax regulations further aggravates this challenge due to the need to keep tax software artifacts accurate and up-to-date.  
These constant changes require continuous revisions and updates to ensure compliance and functionality. Consequently, the current state-of-the-art remains time-consuming and susceptible to errors. 
The authors are supported by an NSF program on  Designing Accountable Software Systems (DASS) to develop principled software engineering tools to improve the accountability of tax preparation software.
In this paper, we discuss key technical challenges in maintaining tax preparation software  by leveraging recent advances in Large Language Models (LLMs).

We posit that the precise and formal language used in tax amendments, as outlined in IRS publications, is amenable to automatic translation into executable software code via LLMs. In addition to natural language processing, LLMs have demonstrated significant potential in generating code~\cite{hindle2016naturalness,fan2023large,li2023starcoder}, thanks to the naturalness of software and the availability of extensive training datasets from software repositories. Our work explores the opportunities and challenges of leveraging LLMs to maintain tax preparation software as it responds to changes in tax laws.

\vspace{0.5em}\noindent\textbf{Testing and Debugging of Tax Prep Software.} The authors in their prior works~\cite{ICSE-SEIS23,srinivas2023potential} focused on the trustworthiness of tax prep software.
One important obstacle to validate the correctness of tax prep software against the tax code, as outlined in the various publications by the IRS, is the \textit{oracle} problem~\cite{6963470}: the class of correctness requirements for tax preparation systems
are not explicitly available since the correct tax-filing is highly subjective to individual taxpayers.
Given the relevant information about an individual, resolving the correct decision for
that individual requires accounting and legal expertise. The authors made a critical observations connecting
the principle of \emph{common law} and \emph{stare decisis} to the metamorphic specifications: 
the correctness of tax preparation software must also be viewed in comparison with similarly situated
taxpayers. One key contribution of these prior works is to explicate formal representations of these properties from the latest Internal Revenue System (IRS) documents (see~\cite{srinivas2023potential} for more info).
In addition, we presented a framework, called \texttt{TenForty}~\cite{ICSE-SEIS23} that automatically generate
test cases from these metamorphic specifications to ensure the trustworthiness of tax prep software.

\vspace{0.5em}\noindent\textbf{LLMs for Maintainability.} 
Following the new tax legislation and the IRS publications of new regulations, the tax prep software needs to be updated to reflect the changes in the software artifacts. However, as the tax law has evolved over different years, updating the corresponding software manually is error-prone and tedious. 
We study the following research question: 
\vspace{-0.4em}
\begin{myboxii}
    \emph{Can we leverage recent breakthroughs in AI, in particular with pre-trained LLMs to assist software developers in automatically updating the implementation of tax law in software artifacts?
} 
\end{myboxii}

LLMs produce a probability distribution over their outputs and thus, are able to generate several candidate solutions with potentially widely differing characteristics.
Our ability to \emph{rank} solutions based on their fitness is a key challenge in employing LLMs for correct software implementations of the tax code. Equipped with a reliable ranking mechanism, one can invoke LLMs in what is known as chain-of-thought reasoning\cite{wei2022chain} to iteratively improve a candidate solution. In this paper, we focus on the problem of \emph{ranking} candidate solutions generated by the LLMs.

\vspace{0.5em}\noindent\textbf{Experimental Setup and Results.} 
For our experiments, we focus on generating functions to compute three key tax calculations: 1) tax brackets, 2) tax deductions, and 3) Earned Income Tax Credits (EITC) through LLMs for the tax year 2021. We use two variants of OpenAI's LLM, ChatGPT (i.e., GPT-4.0 and GPT-3.5), and prompt them with descriptions from the tax publications under two distinct scenarios: 1) with the reference implementation from tax year 2020, and 2) without any reference implementations (direct prompting). In response to these prompts, the LLMs generated 10 candidate code implementations. We first use well-established ranking metrics such as \texttt{CodeBertScore}~\cite{zhou-etal-2023-codebertscore} and compare their ranking outcomes to the ground truth implementations. We found that the existing metrics often fail to rank the candidate implementations in a way that the highest-ranked candidates have the lowest errors compared to the ground truth implementations. Then, we introduce a new metric, \texttt{MajorityVote}, where we take the majority votes of candidates in ranking them (i.e., an implementation that agrees the most with other candidates is considered a high-rank candidate). Our experiments show that a combination of \texttt{CodeBertScore} and \texttt{MajorityVote} outperformed each metric in isolation.

Our results show that when the LLMs are prompted without the reference code of the prior year, the top ranked candidates, generated by GPT-3.5, achieved an accuracy between 0\% to 2\% whereas those by GPT4.0 achieved an accuracy between 43\%-100\%. 
On the other, when prompted with the implementation of tax prep software from the prior year (given as the context to the LLMs), GPT3.5 and GPT4.0 achieved an accuracy between 21\% to 100\% and 48\% to 100\%, respectively. 
Rather than considering the absolute accuracy, we also study the accuracy of our ranking methods with some threshold where a solution within $\delta$\% of ground truth is considered correct. 
We observe that without the prior year code context, GPT3.0's candidates are far off and barely achieved 10\% accuracy under an $\delta$=10\%. GPT4.0's candidates, on the other hand, when prompted without the contexts from the prior years, achieved 100\% accuracy in all cases when the $\delta$ is at least 7\%. 
Interestingly and somehow surprisingly, when prompted with the code context, GPT3.5's candidates achieved a similar or better accuracy, compared to GPT4.0's candidates, when $\delta$ sets to at least 4\%. 

We share our best practices on prompting LLMs to generate implementations of a tax code, giving a code context to the LLMs, and providing the implementations from the prior code. While our current research focuses on a robust ranking system for identifying the most promising candidates from the LLM-generated tax prep software code, the next phase will focus on validating and refinements of (top-ranked) candidates to understand the extend under which the LLMs can be used to update tax prep software and maintain them automatically. For validation, we plan to integrate the ranking system with the metamorphic specifications and testing framework to ensure the correctness of updated code. If the candidate failed over the correctness requirements, Feedback Prompt Generator (FPG) will analyze the specific errors and create targeted prompts to guide the LLM in generating more accurate code in the next iteration.


\begin{figure*}[t]
    \centering
    \includegraphics[width=1.0\textwidth]{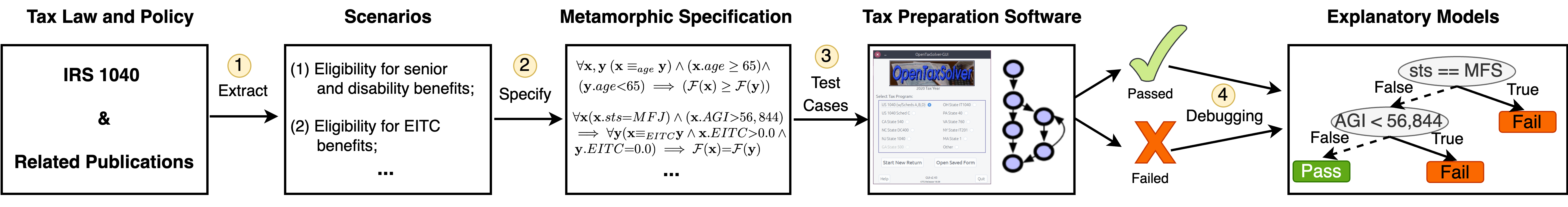}
    \caption{
    {\textsc{TenForty}}: General Framework using Disability and EITC benefits as examples.
    Our approach specifies the correctness requirements from relevant tax policies. Then, it generates random test cases and infers decision trees to localize circumstances under which the software fails to satisfy metamorphic requirements.
    }
    \label{fig:overall-framewok}
\end{figure*}

\section{Maintainability Challenges in Tax Prep Software}
\label{sec:background}
In this section, we briefly review prior work~\cite{ICSE-SEIS23} that leverages metamorphic relations to test the functional correctness of tax preparation software. Using this approach, we have demonstrated how an open-source tax preparation software project failed to correctly update the code to account for new tax legislation. This underscores the importance of automatic methodologies to update tax preparation software.
The key research question in approaching the trustworthiness of tax prep software is the following: 
\begin{myboxii}
\emph{How can one ensure that the tax prep software faithfully implements the tax law code as outlined by various IRS publications such as Form 1040, Publication 596 (EITC), Schedule 8812 (Qualifying Dependents), and Form 8863 (Education Credits)?}
\end{myboxii}

\vspace{0.25 em}
\noindent \textbf{Challenges.} Due to the lack of explicit correctness requirements, one can recourse to pre-existing dataset to test and debug the software. Unfortunately, it is hard to obtain a meaningful labeled dataset—individuals and their “optimal” tax returns—due to obvious privacy and legal concerns. Even when one can learn a good generative model~\cite{mehta2022enhancement} to produce synthetic population, tax software suffers from what is known as \emph{the oracle problem}~\cite{6963470} in software engineering: determining the correct output of an individual decision is time-consuming, expensive, and error-prone due to its highly subjective nature as discussed next. A key observation made in this preliminary work is that a number of compliance specifications can be expressed relating an individual with a counterfactual one. We proposed a formal (first-order) logic (\emph{metamorphic relations}) to express such compliance properties. 

\vspace{0.25 em}
\noindent \textbf{Metamorphic Relations.} We characterized $33$ metamorphic specifications~\cite{srinivas2023potential} from $5$ domains of the U.S. Individual Income Tax Return: (1) Credit for the Elderly or the Disabled~\cite{pub524}, a credit for taxpayers who are aged 65 or older or who are retired on permanent and total disability;
(2) Earned Income Tax Credit (EITC)~\cite{pub596}, a refundable tax credits for lower-income households;
(3) Child Tax Credit (CTC), a nonrefundable credits to reduce the taxes owed based on the number of qualifying children under the age of 17~\cite{8812};
(4) Educational Tax Credit (ETC) that help students with the cost of higher education by lowering their owed taxes or increasing their refund~\cite{8863};
and (5) Itemized Deduction (ID) that is an option for taxpayers with significant tax deductible expenses~\cite{1040sa}. Some examples are: 

\begin{itemize}

\item 
A blind individual must receive similar or better tax benefits when compared to a person without the disability.
This is due to higher standard deductions for blind individuals. This equity specification can be expressed as a metamorphic relation:
\begin{multline*}
\begin{aligned}
\forall {\bf x}, {\bf y} (({\bf x} {\equiv_{blind}} {\bf y}) \wedge ({\bf x}.blind \wedge \neg {\bf y}.blind)) \implies \mathcal{F}({\bf x}) \geq \mathcal{F}({\bf y})
\end{aligned}
\end{multline*}

\item An individual who qualifies for EITC (e.g., income below $56,844$) must receive a higher or equal return than a similar unqualified one.
\begin{equation*}
\begin{aligned}
\forall {\bf x} ({\bf x}.sts{=}MFJ){\implies} & \forall {\bf y} ({\bf x} {\equiv_{AGI}} {\bf y} \wedge {\bf x}.AGI{\leq}56,844 \wedge {\bf y}.AGI{>}56,844) \lor ({\bf x} {\equiv_{L27}} {\bf y} \wedge \\ {\bf x}.L27{>}0.0 \wedge 
& {\bf y}.L27{=}0.0) \lor ({\bf x} {\equiv_{QC}} {\bf y} \wedge {\bf x}.QC{\geq}{\bf y}.QC){\implies}\mathcal{F}({\bf x}){\geq} \mathcal{F}({\bf y})
\end{aligned}
\end{equation*}
\end{itemize}

\vspace{0.25 em}
\noindent \textbf{TenForty Framework.}
We develop an open-source software {\textsc{TenForty}}~\cite{ICSE-SEIS23} (Figure~\ref{fig:overall-framewok})  designed to test and debug tax software.
While it currently focuses on an open-source tax preparation software \textit{OpenTaxSolver}~\cite{openTaxSolver} for the accompanied case study, it can be readily extended to other tax prep software. \textsc{TenForty} allowed us to study the compliance of \textit{OpenTaxSolver}~\cite{openTaxSolver} (tax years of 2018, 2019, 2020, and 2021), a popular open-source tax preparation software~\cite{reddit-opentaxsolver,opensource-opentaxsolver}, in the domains of disability, credits, and deductions that are known to be challenging and error-prone~\cite{IRS-common-mistakes}, leveraging the metamorphic relations. 
{\textsc{TenForty}} generates tens of thousands of random test cases using a given compliance requirements as a metamorphic relation. Furthermore, it explains the circumstances under which the software has failed to comply using an explainable ML model (based on CART decision tree algorithm~\cite{Breiman/1984/CART}). 
Our tool has already revealed three types of failures in \textit{OpenTaxSolver}: missing some eligibility conditions (e.g., married people filing separately status is not eligible to take earned income credits); software fails to satisfy the correctness requirements when the computed tax returns get very close to zero (small non-zero values); and the updated software (e.g., 2021 version updated from 2020 version) that allows users to explicitly opt for an option does not satisfy some correctness requirements in the corner cases.

\section{Overview: Generating Software Code from Tax Code via LLM}
\label{sec:overview}

In this section, we overview the LLM-based code generation and our ranking system using some intuitive examples.
To illustrate the key concepts, we will use simplified examples focusing on snippets of the generated code and the relevant portions of the input context. The full context provided to the LLMs includes detailed instructions, tax policy updates, and, in some cases, the previous year's tax code.  However, for brevity, the figures will only display the code snippets and the contextual elements directly related to those snippets. 

\begin{figure}[!tb]
    \centering
    \includegraphics[width=\textwidth]{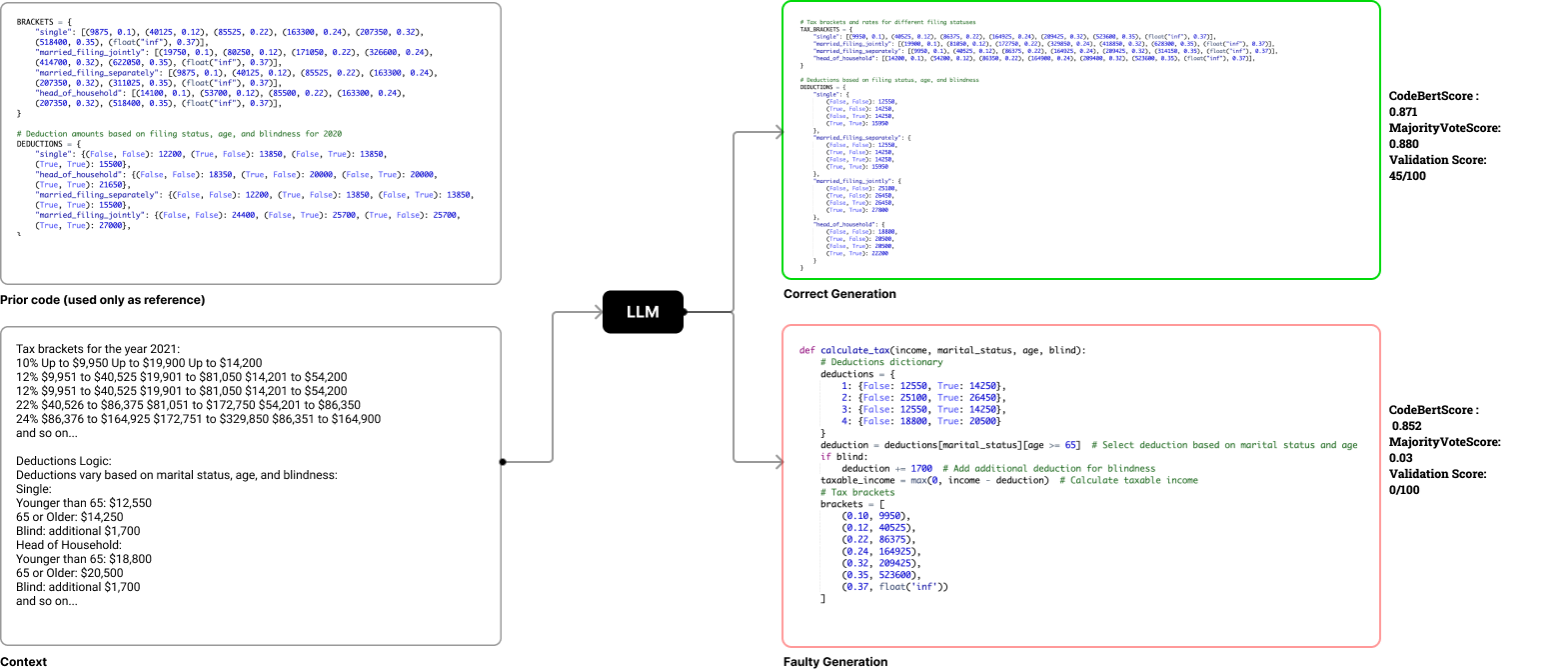}
    \caption{Updating tax brackets without prior software code. Prior code is listed only for clarity to understand CodeBertScore calculation logic; it does not impact the code generation process.} 
    \label{fig:code_gen_no_prior}
\end{figure}

\vspace{0.25 em}
\noindent \textbf{Updating Tax Brackets without Prior Software Code.} Figure~\ref{fig:code_gen_no_prior} presents a visual comparison of two code snippets generated by an LLM when provided with the 2021 tax law updates and a prompt, but without the context of the previous year's code. The code snippet "Faulty Generation" exhibits several flaws, most notably the incomplete definition of the `TAX BRACKETS` dictionary. This error results in a syntactically incorrect program and would likely lead to runtime errors.  The CodeBertScore for this snippet is 0.852, reflecting its lower semantic similarity to the reference code and task instructions due to this significant structural error.

In contrast, the code snippet "Correct Generation" demonstrates a better understanding of the task and the required code structure. 
The `TAX BRACKETS` dictionary is defined correctly, along with the `DEDUCTIONS` dictionary. This snippet is syntactically correct and closer to a functional implementation.  The CodeBertScore for this snippet is 0.871, indicating a much stronger semantic alignment with the reference code and the instructions. The ground truth score (or validation score) is 0.45, reflecting the need for further refinement of this candidate solution.

\begin{figure}[!tbh]
    \centering
    \includegraphics[width=\textwidth]{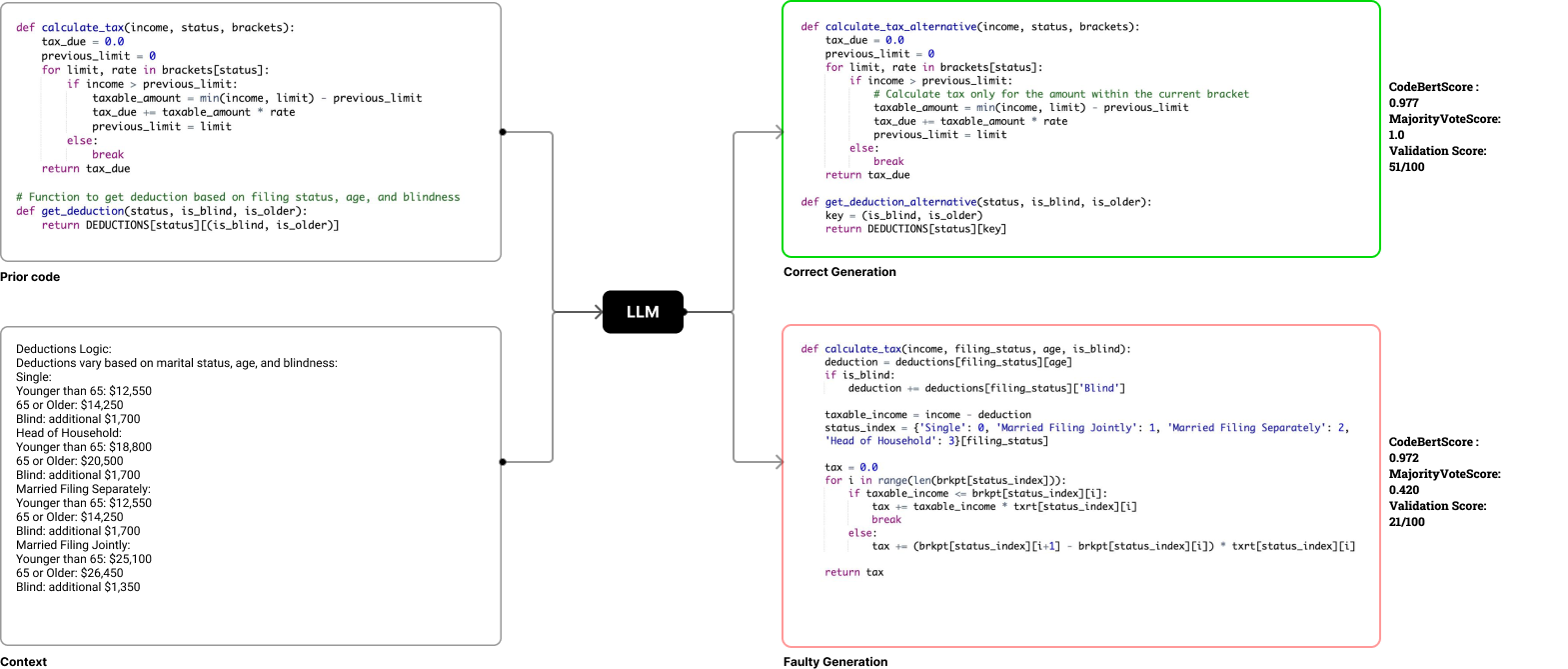}
    \caption{Updating Tax Brackets with Prior Software Code.}
    \label{fig:code_gen_prior}
\end{figure}

\vspace{0.25 em}
\noindent \textbf{Updating Tax Brackets with Prior Software Code.} Figure~\ref{fig:code_gen_prior} displays two code snippets generated by an LLM when provided with the 2021 tax law updates, the 2020 tax code, and a prompt instructing the LLM to update the code. This scenario demonstrates that even with prior code context, LLMs can generate code with logical errors that might not be immediately apparent.

While both code snippets appear structurally similar to the reference code, the snippet "Faulty Generation" contains several logical errors. For instance, there's a potential misalignment between the tax brackets and their corresponding rates, leading to incorrect tax calculations. Additionally, the code incorrectly calculates the blindness deduction by adding a constant value to an already established deduction, potentially causing a double-counting error. These errors would result in incorrect tax outputs for certain inputs, making the code functionally incorrect. Despite these errors, this snippet achieves a CodeBertScore of 0.972, demonstrating that semantic similarity alone is insufficient to guarantee code correctness.

The snippet "Correct Generation", on the other hand, accurately updates the tax calculation logic. It aligns the tax brackets and rates correctly and avoids the double-counting error in the blindness deduction.  This snippet achieves a CodeBertScore of 0.977, slightly higher than the faulty code due to its better semantic alignment.

These comparisons underscore the importance of our multi-faceted ranking approach, which incorporates both CodeBertScore and MajorityVoteScore.  CodeBertScore provides insights into the semantic quality of the generated code, assessing its alignment with the task instructions and reference code (if provided). However, as demonstrated in Figure~\ref{fig:code_gen_prior}, semantic similarity alone is not always sufficient to guarantee functional correctness.  MajorityVoteScore plays a crucial role in detecting logical errors that might not be evident from the code's structure or syntax.  By combining these metrics, our ranking system effectively distinguishes between code candidates with varying levels of quality and correctness, enabling us to select the most promising candidates for further validation and refinement stages of our framework. 

\section{Methodology}

\begin{figure}[t!]
  \centering
  \includegraphics[width=0.9\linewidth]{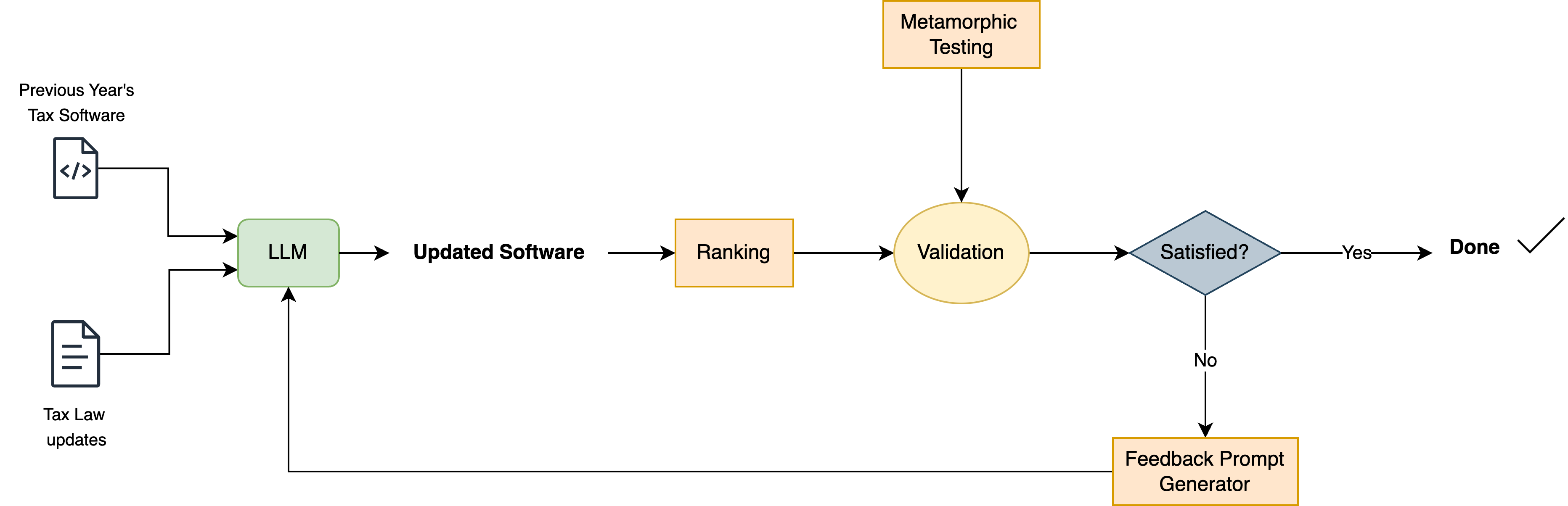}
  \caption{AI-assisted framework to update tax software following the updated tax policies.
  }
  \label{fig:sysarch}
\end{figure}

Figure~\ref{fig:sysarch} illustrates our proposed framework for automatically updating tax preparation software using Large Language Models (LLMs). This framework tackles the challenge of adapting software to the annual revisions in IRS tax policies, aiming to reduce manual effort and increase the trustworthiness. The framework operates as a cyclical process consisting of several key stages:

\begin{enumerate}
\item \textbf{Input and Analysis:}  The process begins by providing the LLM with two essential inputs:
    \begin{itemize}
        \item \textbf{Previous Year's Tax Software Code:} The source code of the existing tax software serves as the base for the update. To understand its importance, 
        we perform experiments without including the previous year's code.  
        \item \textbf{Latest Tax Policy Updates:}  The LLM receives the official IRS publications detailing the changes in tax laws for the current year. 
    \end{itemize}
    
\item \textbf{Code Generation:}  Leveraging the provided inputs, the LLM generates multiple candidate versions of the updated tax software.  The LLM is guided by a series of prompts that provide context, instructions, and previous year's tax calculation code to maximize the alignments of generated code to the desired functionality.
We experiment with two LLMs, ChatGPT3.5 and GPT4.0 to explore their effectiveness in this code generation task.

\item \textbf{Ranking and Selection:} Since the LLMs can generate a large number of candidate code, \textit{the main focus of this paper is to come up with a ranking criteria to identify the most promising candidates.}
We consider the following ranking mechanisms: 

    \begin{itemize}
        \item \textbf{CodeBertScore:} We leverage CodeBERT~\cite{zhou-etal-2023-codebertscore}, a pre-trained model specializing in understanding code, to assess the semantic similarity of the generated code. This metric calculates the cosine similarity between the generated code and both the reference code from the previous year and the IRS policy updates. A higher CodeBertScore indicates a stronger alignment between the generated code and the intended meaning and structure expressed in the reference code and the new tax regulations. 
        
        \item \textbf{MajorityVoteScore:}  We execute the each candidate code with a set of random inputs to quantify the majority vote. These random input profiles cover a diverse range of income levels, filing statuses, and other relevant parameters.
        For each input profile, we run all generated code versions and record their outputs. We then determine the most frequent output across all versions, assuming this "majority vote" output to be the correct answer. The majority vote score of each code version is then calculated as the percentage of inputs for which its output matches the majority vote output.
        
        \item \textbf{WeightedScore:} To determine the overall ranking of the generated code candidates, we employ a weighted average that combines both CodeBertScore and MajorityVoteScore. This approach allows us to prioritize candidates that excel in two key aspects: semantic similarity to the task instructions and reference code (CodeBertScore) and functional correctness in producing accurate tax calculations (MajorityVoteScore). We perform various experiments and found that assigning a weight of 0.6 to CodeBertScore and 0.4 to the MajorityVoteScore works well in practice. The setup also depends on the capabilities of LLMs. More capable LLMs (like GPT-4) consistently generate high-quality code, making the majority vote score a reliable indicator of correctness whereas less capable LLMs (such as GPT-3.5) may exhibit greater inconsistency in code quality, relying too heavily on the majority vote score could lead to misinterpretations.

    \end{itemize}

\item \textbf{Metamorphic Testing:} To further validate the top-ranked code candidates, we employ metamorphic testing~\cite{ICSE-SEIS23,srinivas2023potential}. We previously leverage metamorphic specification and testing to validate the correctness of tax prep software (see Background Section~\ref{sec:background}). After we choose a top-ranked candidate, we use the metamorphic testing paradigm to validate its correctness or obtain failed test-cases to guide a refinement process.

\item \textbf{Feedback Loop:} 
    \begin{itemize}
        \item \textbf{Success:} If a code candidate successfully passes the metamorphic testing stage without any failures, we deem it correct and return the solution to the tax software developers as the correct updated software. 
    
        \item \textbf{Refinement:} In cases where the code fails one or more metamorphic test cases, our framework initiates a feedback loop for iterative refinement. The \textbf{Feedback Prompt Generator (FPG)} analyzes the specific test failures and generates targeted prompts to guide the LLM in rectifying the identified issues.  
    \end{itemize}
\item \textbf{Iteration:} The process of code generation, ranking, metamorphic testing, and feedback-driven refinement may iterates multiple times. This process successes if the generated software successfully passes all metamorphic tests, and we obtain some statistically or formal guarantees on the correctness. 
\end{enumerate}

Overall, the framework~\ref{fig:overall-framewok} provides a means to update and maintain tax prep software automatically via LLMs. 
\textit{This paper only focuses on the ranking systems for the LLM-generated code and discuss the technical challenges in using LLMs to update tax prep software as the tax law changes each year.}
While our prior works used metamorphic testing~\cite{ICSE-SEIS23,srinivas2023potential} to ensure the correctness of general tax prep software, more work is needed to integrate it as the validation component in the framework~\ref{fig:overall-framewok}. Also, the feedback prompt generators may not be trivial and require extensive future works to guide LLMs in generating candidate code.

\section{Experiments and Results}
\label{sec:experiments}

\subsection{Updating tax prep software without prior code context via LLMs}
This section explores the performance of LLMs in updating tax preparation software when \textit{no} context about the previous year's code is provided. This scenario examines whether the LLMs are capable in generating tax prep software code from scratch.

\vspace{0.25 em}
\noindent \textbf{Procedure.}
We follow the same general framework outlined in methodology section, but omit the initial input of the previous year's code. The LLM receives only the following:

\begin{itemize}
    \item \textit{Tax Policy Updates:} The official IRS publications describing the changes in tax laws for the current year (2021 in our experiments).
    \item \textit{Prompt Engineering:} A set of instructions guiding the LLM to generate the updated software. 
\end{itemize}

The LLM then generates multiple candidate code versions. These versions are ranked using the CodeBertScore and majority vote accuracy metrics. The top-ranked candidates undergo metamorphic testing to validate their correctness.

\vspace{0.25 em}
\noindent \textbf{Prompt Engineering.}
Here's a specific prompt used to guide the LLM in generating code for the "Brackets Only" scenario:

\begin{quote}
\textit{Objective}: Develop a Python script to calculate federal income tax for the year 2021. The script should accurately compute tax based on the user's annual income and marital status, incorporating the 2021 tax brackets.

\textit{Data Structures}:
\begin{itemize}
    \item Use dictionaries to map tax brackets for different filing statuses (Single, Married Filing Jointly, Married Filing Separately, Head of Household).
    \item Ensure keys are accurately used to prevent KeyError and validate their presence before access.
\end{itemize}

\textit{User Inputs}:
\begin{verbatim}
  - `income`: Collect as a float using input(), representing 
   the user's annual income in USD.
  - `marital_status`: Integer (1-4); 1=Single, 2=Married Filing Jointly, 
   3=Married Filing Separately, 4=Head of Household.
\end{verbatim}

\textit{Requirements}:
\begin{itemize}
    \item The script must compute the tax using the provided tax brackets.
    \item Output the tax amount in dollars formatted to two decimals (e.g., print(f"Tax amount: \${tax:.2f}")).
    \item Include error handling for user inputs to ensure they are within valid ranges and formats.
\end{itemize}

[2021 Tax Brackets (concrete numbers should be provided)]
\end{quote}

\vspace{0.25 em}
\noindent \textbf{General Prompt Template.}
We adapt the following template for different scenarios, modifying the specific instructions and data as needed:

\begin{quote}
\textit{Objective}: [Clearly state the task, e.g., "Develop a Python script to calculate federal income tax for the year 2021."]

\textit{Data Structures}: [Specify the expected data structures, e.g., dictionaries for tax brackets and deductions.]

\textit{User Inputs}: [List the required user inputs and their data types.]

\textit{Requirements}: [Outline the functional requirements of the code, e.g., tax calculation logic, output format, error handling.]

[Provide any relevant tax policy data, e.g., tax brackets, deduction amounts, EITC rules.]
\end{quote}

\begin{table}[]
\caption{Results for top 4 ranked code generations out of 10 without prior code.\stsays{}}
\resizebox{1.0\textwidth}{!}{%
\begin{tabular}{|l|l|l|r|r|r|l|}
\hline
\textbf{Scenarios}                            & \textbf{LLM}                                    & \textbf{Versions}                  & \multicolumn{1}{l|}{\textbf{CodeBertScore}} & \multicolumn{1}{l|}{\textbf{MajorityVoteScore}} & \multicolumn{1}{l|}{\textbf{WeightedScore}} & \cellcolor[HTML]{FFFFFF}\textbf{Ground Truth Score} \\ \hline
\multicolumn{1}{|c|}{}                                    & \cellcolor[HTML]{C0C0C0}                        & \cellcolor[HTML]{C0C0C0}Version 7  & \cellcolor[HTML]{C0C0C0}0.9            & \cellcolor[HTML]{C0C0C0}1                   & \cellcolor[HTML]{C0C0C0}0.935                & \cellcolor[HTML]{C0C0C0}100/100                     \\ \cline{3-7} 
\multicolumn{1}{|c|}{}                                    & \cellcolor[HTML]{C0C0C0}                        & \cellcolor[HTML]{C0C0C0}Version 2  & \cellcolor[HTML]{C0C0C0}0.899          & \cellcolor[HTML]{C0C0C0}1                   & \cellcolor[HTML]{C0C0C0}0.934                & \cellcolor[HTML]{C0C0C0}100/100                     \\ \cline{3-7} 
\multicolumn{1}{|c|}{}                                    & \cellcolor[HTML]{C0C0C0}                        & \cellcolor[HTML]{C0C0C0}Version 4  & \cellcolor[HTML]{C0C0C0}0.899          & \cellcolor[HTML]{C0C0C0}1                   & \cellcolor[HTML]{C0C0C0}0.934                & \cellcolor[HTML]{C0C0C0}100/100                     \\ \cline{3-7} 
\multicolumn{1}{|c|}{}                                    & \multirow{-4}{*}{\cellcolor[HTML]{C0C0C0}GPT 4} & \cellcolor[HTML]{C0C0C0}Version 5  & \cellcolor[HTML]{C0C0C0}0.899          & \cellcolor[HTML]{C0C0C0}1                   & \cellcolor[HTML]{C0C0C0}0.934                & \cellcolor[HTML]{C0C0C0}100/100                     \\ \cline{2-7} 
\multicolumn{1}{|c|}{}                                    &                                                 & Version 9                          & 0.894                                  & 0.94                                        & 0.91                                         & 0/100                                               \\ \cline{3-7} 
\multicolumn{1}{|c|}{}                                    &                                                 & Version 2                          & 0.892                                  & 0.94                                        & 0.909                                        & 0/100                                               \\ \cline{3-7} 
\multicolumn{1}{|c|}{}                                    &                                                 & Version 6                          & 0.903                                  & 0.06                                        & 0.608                                        & 0/100                                               \\ \cline{3-7} 
\multicolumn{1}{|c|}{\multirow{-8}{*}{\textit{Brackets}}} & \multirow{-4}{*}{GPT 3.5}                       & Version 8                          & 0.894                                  & 0.06                                        & 0.602                                        & 0/100                                               \\ \hline
                                                          & \cellcolor[HTML]{C0C0C0}                        & \cellcolor[HTML]{C0C0C0}Version 3  & \cellcolor[HTML]{C0C0C0}0.871          & \cellcolor[HTML]{C0C0C0}0.88                & \cellcolor[HTML]{C0C0C0}0.875                & \cellcolor[HTML]{C0C0C0}45/100                      \\ \cline{3-7} 
                                                          & \cellcolor[HTML]{C0C0C0}                        & \cellcolor[HTML]{C0C0C0}Version 2  & \cellcolor[HTML]{C0C0C0}0.866          & \cellcolor[HTML]{C0C0C0}0.88                & \cellcolor[HTML]{C0C0C0}0.871                & \cellcolor[HTML]{C0C0C0}45/100                      \\ \cline{3-7} 
                                                          & \cellcolor[HTML]{C0C0C0}                        & \cellcolor[HTML]{C0C0C0}Version 5  & \cellcolor[HTML]{C0C0C0}0.861          & \cellcolor[HTML]{C0C0C0}0.88                & \cellcolor[HTML]{C0C0C0}0.869                & \cellcolor[HTML]{C0C0C0}45/100                      \\ \cline{3-7} 
                                                          & \multirow{-4}{*}{\cellcolor[HTML]{C0C0C0}GPT 4} & \cellcolor[HTML]{C0C0C0}Version 10 & \cellcolor[HTML]{C0C0C0}0.887          & \cellcolor[HTML]{C0C0C0}0.12                & \cellcolor[HTML]{C0C0C0}0.58                 & \cellcolor[HTML]{C0C0C0}0/100                       \\ \cline{2-7} 
                                                          &                                                 & Version 2                          & 0.859                                  & 1                                           & 0.916                                        & 1/100                                               \\ \cline{3-7} 
                                                          &                                                 & Version 1                          & 0.859                                  & 1                                           & 0.916                                        & 1/100                                               \\ \cline{3-7} 
                                                          &                                                 & Version 6                          & 0.858                                  & 1                                           & 0.915                                        & 1/100                                               \\ \cline{3-7} 
\multirow{-8}{*}{\textit{Brackets + Deductions}}          & \multirow{-4}{*}{GPT 3.5}                       & Version 10                         & 0.858                                  & 1                                           & 0.915                                        & 1/100                                               \\ \hline
                                                          & \cellcolor[HTML]{C0C0C0}                        & \cellcolor[HTML]{C0C0C0}Version 7  & \cellcolor[HTML]{C0C0C0}0.883          & \cellcolor[HTML]{C0C0C0}0.79                & \cellcolor[HTML]{C0C0C0}0.827                & \cellcolor[HTML]{C0C0C0}43/100                      \\ \cline{3-7} 
                                                          & \cellcolor[HTML]{C0C0C0}                        & \cellcolor[HTML]{C0C0C0}Version 1  & \cellcolor[HTML]{C0C0C0}0.863          & \cellcolor[HTML]{C0C0C0}0.7                 & \cellcolor[HTML]{C0C0C0}0.765                & \cellcolor[HTML]{C0C0C0}25/100                      \\ \cline{3-7} 
                                                          & \cellcolor[HTML]{C0C0C0}                        & \cellcolor[HTML]{C0C0C0}Version 5  & \cellcolor[HTML]{C0C0C0}0.87           & \cellcolor[HTML]{C0C0C0}0.61                & \cellcolor[HTML]{C0C0C0}0.714                & \cellcolor[HTML]{C0C0C0}32/100                      \\ \cline{3-7} 
                                                          & \multirow{-4}{*}{\cellcolor[HTML]{C0C0C0}GPT 4} & \cellcolor[HTML]{C0C0C0}Version 6  & \cellcolor[HTML]{C0C0C0}0.857          & \cellcolor[HTML]{C0C0C0}0.61                & \cellcolor[HTML]{C0C0C0}0.709                & \cellcolor[HTML]{C0C0C0}32/100                      \\ \cline{2-7} 
                                                          &                                                 & Version 6                          & 0.852                                  & 1                                           & 0.941                                        & 2/100                                               \\ \cline{3-7} 
                                                          &                                                 & Version 2                          & 0.851                                  & 0.98                                        & 0.929                                        & 0/100                                               \\ \cline{3-7} 
                                                          &                                                 & Version 10                         & 0.845                                  & 0.98                                        & 0.926                                        & 0/100                                               \\ \cline{3-7} 
\multirow{-8}{*}{\textit{Brackets+Ded+EITC}}              & \multirow{-4}{*}{GPT 3.5}                       & Version 3                          & 0.845                                  & 0.5                                         & 0.638                                        & 0/100                                               \\ \hline
\end{tabular}
\label{table1}
}
\end{table}

\begin{figure}[!tbh]
    \centering
    \begin{subfigure}{0.32\textwidth}
        \includegraphics[width=\linewidth]{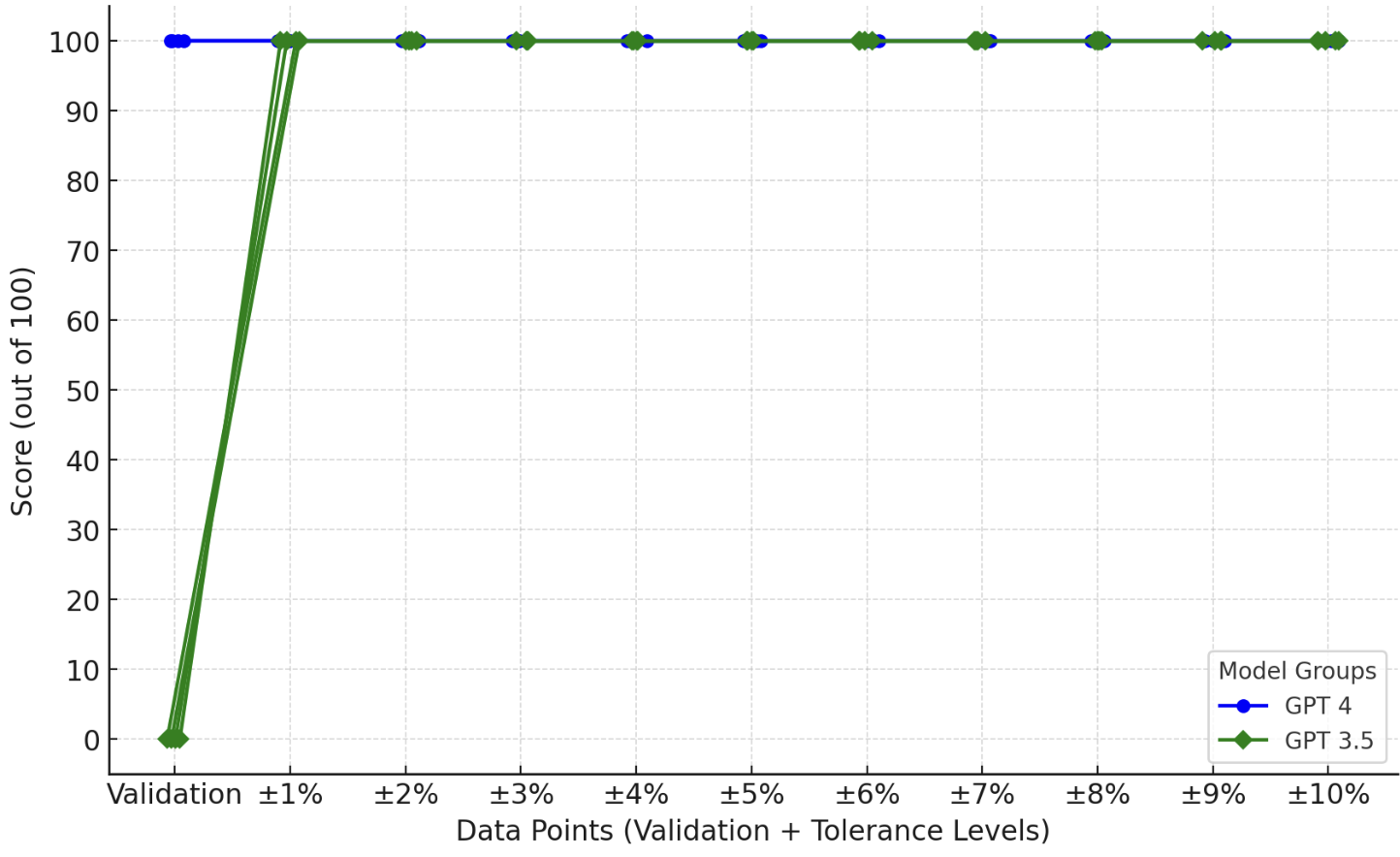}
        \caption{Brackets}
    \end{subfigure}
    \hfill
    \begin{subfigure}{0.32\textwidth}
        \includegraphics[width=\linewidth]{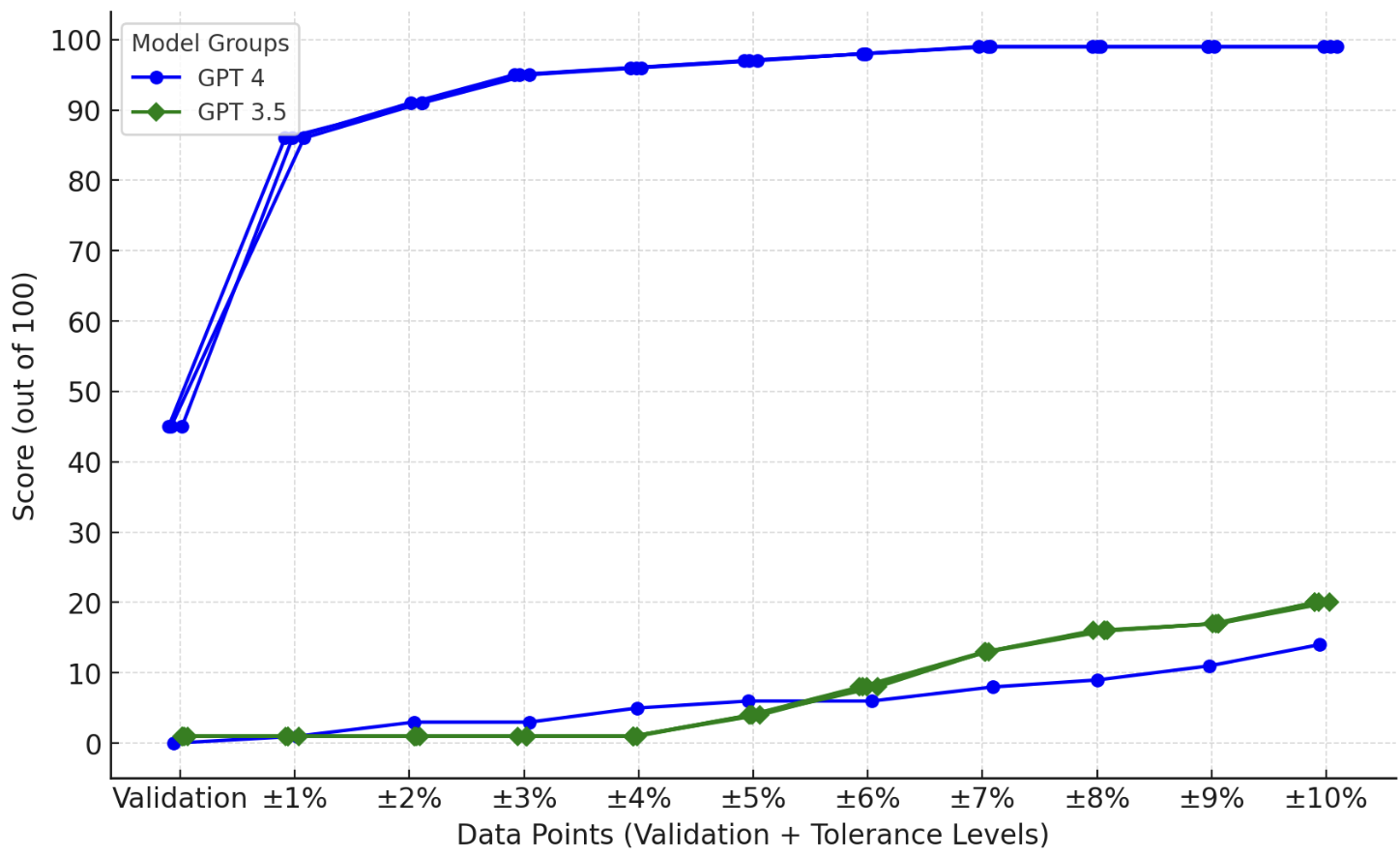}
        \caption{Brackets+Deductions}
    \end{subfigure}
    \hfill
    \begin{subfigure}{0.32\textwidth}
        \includegraphics[width=\linewidth]{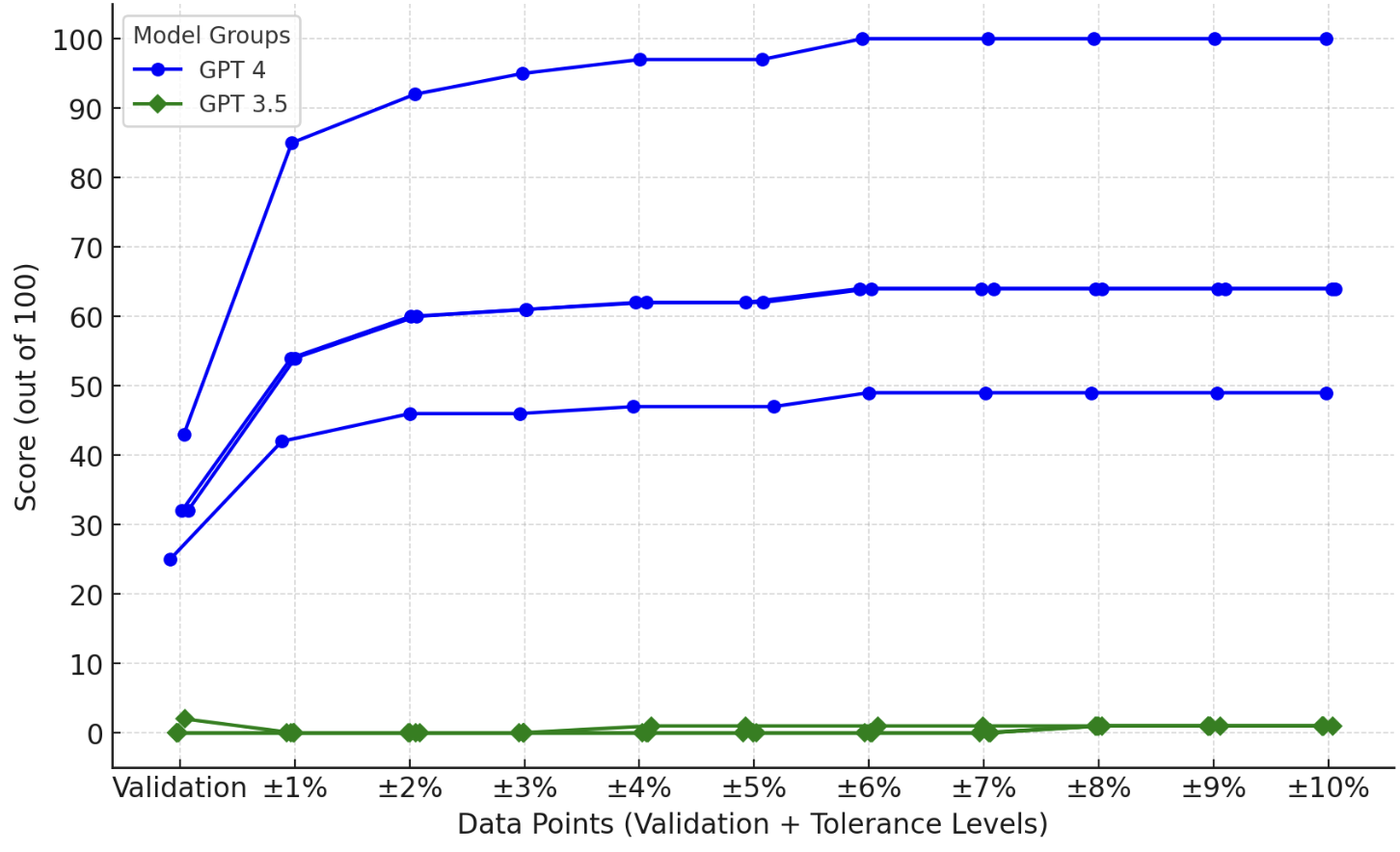}
        \caption{Brackets+Deductions+EITC}
    \end{subfigure}
    \caption{Scenarios without prior code for 4 top ranked candidates per ChatGPT-3.5/4.0. }
\label{fig:accuracy-vs-tolerance-1}
\end{figure}

\vspace{0.25 em}
\noindent \textbf{Results and Discussion.}
Table~\ref{table1} presents the results for the top-performing code candidates generated by GPT-4 and GPT-3.5 in each scenario without prior code context.  As shown in the table, LLMs demonstrate a varied level of performance depending on the complexity of the scenario and the specific LLM model used.

\begin{itemize}
    \item \textbf{Overall Lower Performance:}  We observe a general trend of lower performance across all scenarios when the LLM is not provided with the previous year's code. Both GPT-4 and GPT-3.5 exhibit lower MajorityVoteScore and CodeBertScore compared to when they have the reference code for guidance. Because the LLM has to come up with the logic of the code itself as opposed to when they are presented with previous code and can use it as guidance.
    
    \item \textbf{GPT-4's Continued Superiority:} Despite the lack of prior code, GPT-4 consistently outperforms GPT-3.5. This suggests that GPT-4 possesses a stronger capacity for understanding instructions and generating correct code from scratch.
    
    \item \textbf{Accuracy Decline with Complexity:}  As the scenarios become more complex with the inclusion of deductions and EITC, the accuracy of both LLMs drops noticeably, particularly for GPT-3.5. This underscores the challenges LLMs face in generating intricate tax logic from scratch without the benefit of a reference code to guide the process.
    
    \item \textbf{MajorityVoteScore vs. Ground Truth Discrepancies:}  A crucial observation is the occasional disparity between high majority vote accuracy and significantly lower ground truth matching scores. This discrepancy suggests that LLMs can sometimes generate code that consistently produces the most common output, but still contains subtle errors that cause deviations from the ideal calculation. This finding emphasizes the importance of considering other ranking metrics and their combinations. 
\end{itemize}

While Table~\ref{table1} presents the absolute accuracy of the generated code candidates, it doesn't provide insights into how close their outputs are to the true tax calculations. To gain a more nuanced understanding of the code's correctness, we analyze the accuracy with respect to acceptable error margins from the ground truth.
Furthermore, the scatter plots in Figure \ref{fig:accuracy-vs-tolerance-1} illustrate the accuracy of generated code candidates (y-axis) based on an acceptable threshold of error margins from the ground truth. 
The plots show the consistency of ChatGPT-4.0 compared to ChatGPT-3.5. Even if the outputs of code is not the exact ground truth, the output from ChatGPT-4.0 generated codes is almost always within the 10\% error margins of ground truth values which suggests that the logic of the code is sound but there might be some small problems. This also emphasizes that the ranking part of our framework works well in finding codes that have the potential for fixing. ChatGPT-3.5, on the other hand, shows that it cannot generate high quality and sounds codes from scratch consistently. ChatGPT-3.5 is especially fragile when it does not generate at least 2 good codes that can have consensus on outputs. While Table  presents the absolute accuracy of the generated code candidates when LLMs are not provided with prior year code, it does not reveal how close their outputs are to the true tax calculations. To understand the proximity of generated outputs to the ground truth, we analyze the accuracy within an acceptable error margins. 

When prompted without the reference code, GPT-3.5's top-ranked candidates struggle to achieve high accuracy, even with a generous error margin. They barely reach 10\% accuracy even when allowing a $\delta$ of 10\%. This suggests that GPT-3.5, when generating code from scratch, often produces codes that may have a wrong logic.
Conversely, GPT-4.0's top-ranked candidates, even without prior code context, exhibit better performance.  They consistently achieve 100\% accuracy when considering a $\delta$ threshold of at least 7\%. 
However, it's important to note that achieving perfect accuracy at a 7\% error margin still indicates the presence of errors that require refinement.

\subsection{Updating tax prep software with prior code context via LLMs}
This section investigates the performance of LLMs in updating tax software when provided with the previous year's code as context. This scenario emulates a more realistic use case where the LLM can leverage existing code structure and logic as a foundation for incorporating tax policy changes.

\vspace{0.25 em}
\noindent \textbf{Procedure.}
Following the framework outlined in the methodology section, the LLM receives the following inputs:

\begin{itemize}
    \item \textit{Previous Year's Tax Software Code:} The source code of the existing tax software (for 2020 in our experiments) acts as a basis for the update.
    \item \textit{Tax Policy Updates:} The official IRS publications detailing the changes in tax laws for the current year (2021 in our case).
    \item \textit{Prompt:} A set of instructions that guide the LLM in modifying the provided code to reflect the new tax policy. 
\end{itemize}

The LLM generates multiple updated code versions, which are then ranked using CodeBertScore and MajorityVoteScore. 

\vspace{0.25 em}
\noindent \textbf{Prompt Engineering.}
Here's a specific prompt used to guide the LLM in generating code for the "Brackets+Deductions" scenario:

\begin{quote}
\textit{Objective:} Update the provided Python code to calculate federal income tax for the year 2021. The updated script should accurately compute tax based on the user's annual income, marital status, age, and blindness status, incorporating the 2021 tax brackets and standard deductions.

\textit{Reference Python Code (2020):}
\vspace{-0.5 em}
\begin{verbatim}
  # Constants for tax brackets and rates for 2020
   BRACKETS = {
      "single": [(9875, 0.1), (40125, 0.12), (85525, 0.22), 
                 (163300, 0.24), (207350, 0.32), (518400, 0.35), 
                 (float("inf"), 0.37)],
      # ... [Other filing statuses] ...
   }
 
  # Deduction amounts based on filing status, age, and blindness for 2020
   DEDUCTIONS = {
      "single": {(False, False): 12200, (True, False): 13850, 
                 (False, True): 13850, (True, True): 15500},
      # ... [Other filing statuses] ...
   }

  # ... [Rest of the 2020 code] ...
\end{verbatim}

\textit{Instructions (User Inputs + Requirements):}
\begin{itemize}
    \item Update the \texttt{BRACKETS} dictionary to reflect the 2021 tax brackets.
    \item Update the \texttt{DEDUCTIONS} dictionary to incorporate the 2021 standard deduction.
    \item Ensure the script accurately calculates tax based on income, filing status, age, and blindness status.
    \item Maintain the same user input format (income, marital status, age, blindness).
    \item Output the tax amount in dollars formatted to two decimals. 
\end{itemize}

[The 2021 tax brackets and deduction amounts.]
\end{quote}

\vspace{0.25 em}
\noindent \textbf{General Prompt Template with Code Context:}
\begin{quote}
\textit{Objective:} [Clearly state the task, including the year of the provided code and the desired year for the updated code.]

\textit{Reference Python Code:} \texttt{[Previous year's code.]}

\textit{User Inputs:} [Provide specific instructions on how to update the provided code, referencing variable names or functions as needed.]

\textit{Requirements:} [Specify any changes in user input format or output requirements.]

[Provide the necessary tax policy data for the target year.]
\end{quote}

\begin{table}[]
\caption{ Results for top 4 ranked code generations out of 10 with prior code.}
\resizebox{1.0\textwidth}{!}{%
\begin{tabular}{|l|l|l|r|r|r|l|}
\hline
\textbf{Scenario}                                         & \textbf{LLM}                                    & \textbf{Versions}              & \multicolumn{1}{l|}{\textbf{CodeBertScore}} & \multicolumn{1}{l|}{\textbf{MajorityVoteScore}} & \multicolumn{1}{l|}{\textbf{WeightedScore}} & \cellcolor[HTML]{FFFFFF}\textbf{Ground Truth Score} \\ \hline
\multicolumn{1}{|c|}{}                                    & \cellcolor[HTML]{C0C0C0}                        & \cellcolor[HTML]{C0C0C0}Version 3  & \cellcolor[HTML]{C0C0C0}0.914          & \cellcolor[HTML]{C0C0C0}1                   & \cellcolor[HTML]{C0C0C0}0.944                & \cellcolor[HTML]{C0C0C0}100/100                     \\ \cline{3-7} 
\multicolumn{1}{|c|}{}                                    & \cellcolor[HTML]{C0C0C0}                        & \cellcolor[HTML]{C0C0C0}Version 5  & \cellcolor[HTML]{C0C0C0}0.911          & \cellcolor[HTML]{C0C0C0}1                   & \cellcolor[HTML]{C0C0C0}0.942                & \cellcolor[HTML]{C0C0C0}100/100                     \\ \cline{3-7} 
\multicolumn{1}{|c|}{}                                    & \cellcolor[HTML]{C0C0C0}                        & \cellcolor[HTML]{C0C0C0}Version 9  & \cellcolor[HTML]{C0C0C0}0.911          & \cellcolor[HTML]{C0C0C0}1                   & \cellcolor[HTML]{C0C0C0}0.592                & \cellcolor[HTML]{C0C0C0}100/100                     \\ \cline{3-7} 
\multicolumn{1}{|c|}{}                                    & \multirow{-4}{*}{\cellcolor[HTML]{C0C0C0}GPT 4} & \cellcolor[HTML]{C0C0C0}Version 4  & \cellcolor[HTML]{C0C0C0}0.91           & \cellcolor[HTML]{C0C0C0}1                   & \cellcolor[HTML]{C0C0C0}0.941                & \cellcolor[HTML]{C0C0C0}100/100                     \\ \cline{2-7} 
\multicolumn{1}{|c|}{}                                    &                                                 & Version 1                          & 0.941                                  & 1                                           & 0.962                                        & 100/100                                             \\ \cline{3-7} 
\multicolumn{1}{|c|}{}                                    &                                                 & Version 2                          & 0.939                                  & 1                                           & 0.96                                         & 100/100                                             \\ \cline{3-7} 
\multicolumn{1}{|c|}{}                                    &                                                 & Version 7                          & 0.937                                  & 1                                           & 0.959                                        & 100/100                                             \\ \cline{3-7} 
\multicolumn{1}{|c|}{\multirow{-8}{*}{\textit{Brackets}}} & \multirow{-4}{*}{GPT 3.5}                       & Version 8                          & 0.936                                  & 0.59                                        & 0.815                                        & 59/100                                              \\ \hline
                                                          & \cellcolor[HTML]{C0C0C0}                        & \cellcolor[HTML]{C0C0C0}Version 7  & \cellcolor[HTML]{C0C0C0}0.972          & \cellcolor[HTML]{C0C0C0}1                   & \cellcolor[HTML]{C0C0C0}0.983                & \cellcolor[HTML]{C0C0C0}51/100                      \\ \cline{3-7} 
                                                          & \cellcolor[HTML]{C0C0C0}                        & \cellcolor[HTML]{C0C0C0}Version 5  & \cellcolor[HTML]{C0C0C0}0.972          & \cellcolor[HTML]{C0C0C0}1                   & \cellcolor[HTML]{C0C0C0}0.983                & \cellcolor[HTML]{C0C0C0}51/100                      \\ \cline{3-7} 
                                                          & \cellcolor[HTML]{C0C0C0}                        & \cellcolor[HTML]{C0C0C0}Version 3  & \cellcolor[HTML]{C0C0C0}0.972          & \cellcolor[HTML]{C0C0C0}1                   & \cellcolor[HTML]{C0C0C0}0.983                & \cellcolor[HTML]{C0C0C0}51/100                      \\ \cline{3-7} 
                                                          & \multirow{-4}{*}{\cellcolor[HTML]{C0C0C0}GPT 4} & \cellcolor[HTML]{C0C0C0}Version 6  & \cellcolor[HTML]{C0C0C0}0.972          & \cellcolor[HTML]{C0C0C0}1                   & \cellcolor[HTML]{C0C0C0}0.983                & \cellcolor[HTML]{C0C0C0}51/100                      \\ \cline{2-7} 
                                                          &                                                 & Version 4                          & 0.976                                  & 1                                           & 0.99                                         & 21/100                                              \\ \cline{3-7} 
                                                          &                                                 & Version 3                          & 0.976                                  & 1                                           & 0.99                                         & 21/100                                              \\ \cline{3-7} 
                                                          &                                                 & Version 6                          & 0.975                                  & 1                                           & 0.99                                         & 21/100                                              \\ \cline{3-7} 
\multirow{-8}{*}{\textit{Brackets + Deductions}}          & \multirow{-4}{*}{GPT 3.5}                       & Version 5                          & 0.975                                  & 1                                           & 0.99                                         & 21/100                                              \\ \hline
                                                          & \cellcolor[HTML]{C0C0C0}                        & \cellcolor[HTML]{C0C0C0}Version 6  & \cellcolor[HTML]{C0C0C0}0.978          & \cellcolor[HTML]{C0C0C0}1                   & \cellcolor[HTML]{C0C0C0}0.991                & \cellcolor[HTML]{C0C0C0}48/100                      \\ \cline{3-7} 
                                                          & \cellcolor[HTML]{C0C0C0}                        & \cellcolor[HTML]{C0C0C0}Version 8  & \cellcolor[HTML]{C0C0C0}0.978          & \cellcolor[HTML]{C0C0C0}1                   & \cellcolor[HTML]{C0C0C0}0.991                & \cellcolor[HTML]{C0C0C0}48/100                      \\ \cline{3-7} 
                                                          & \cellcolor[HTML]{C0C0C0}                        & \cellcolor[HTML]{C0C0C0}Version 10 & \cellcolor[HTML]{C0C0C0}0.976          & \cellcolor[HTML]{C0C0C0}1                   & \cellcolor[HTML]{C0C0C0}0.991                & \cellcolor[HTML]{C0C0C0}48/100                      \\ \cline{3-7} 
                                                          & \multirow{-4}{*}{\cellcolor[HTML]{C0C0C0}GPT 4} & \cellcolor[HTML]{C0C0C0}Version 3  & \cellcolor[HTML]{C0C0C0}0.976          & \cellcolor[HTML]{C0C0C0}1                   & \cellcolor[HTML]{C0C0C0}0.991                & \cellcolor[HTML]{C0C0C0}48/100                      \\ \cline{2-7} 
                                                          &                                                 & Version 1                          & 0.986                                  & 1                                           & 0.994                                        & 56/100                                              \\ \cline{3-7} 
                                                          &                                                 & Version 2                          & 0.977                                  & 0.92                                        & 0.943                                        & 56/100                                              \\ \cline{3-7} 
                                                          &                                                 & Version 7                          & 0.977                                  & 0.56                                        & 0.727                                        & 35/100                                              \\ \cline{3-7} 
\multirow{-8}{*}{\textit{Brackets+Ded+EITC}}              & \multirow{-4}{*}{GPT 3.5}                       & Version 3                          & 0.977                                  & 0.56                                        & 0.727                                        & 35/100                                              \\ \hline
\end{tabular}
\label{table2}
}
\end{table}

\begin{figure}[!t]
    \centering
    \begin{subfigure}{0.32\textwidth}
        \includegraphics[width=\linewidth]{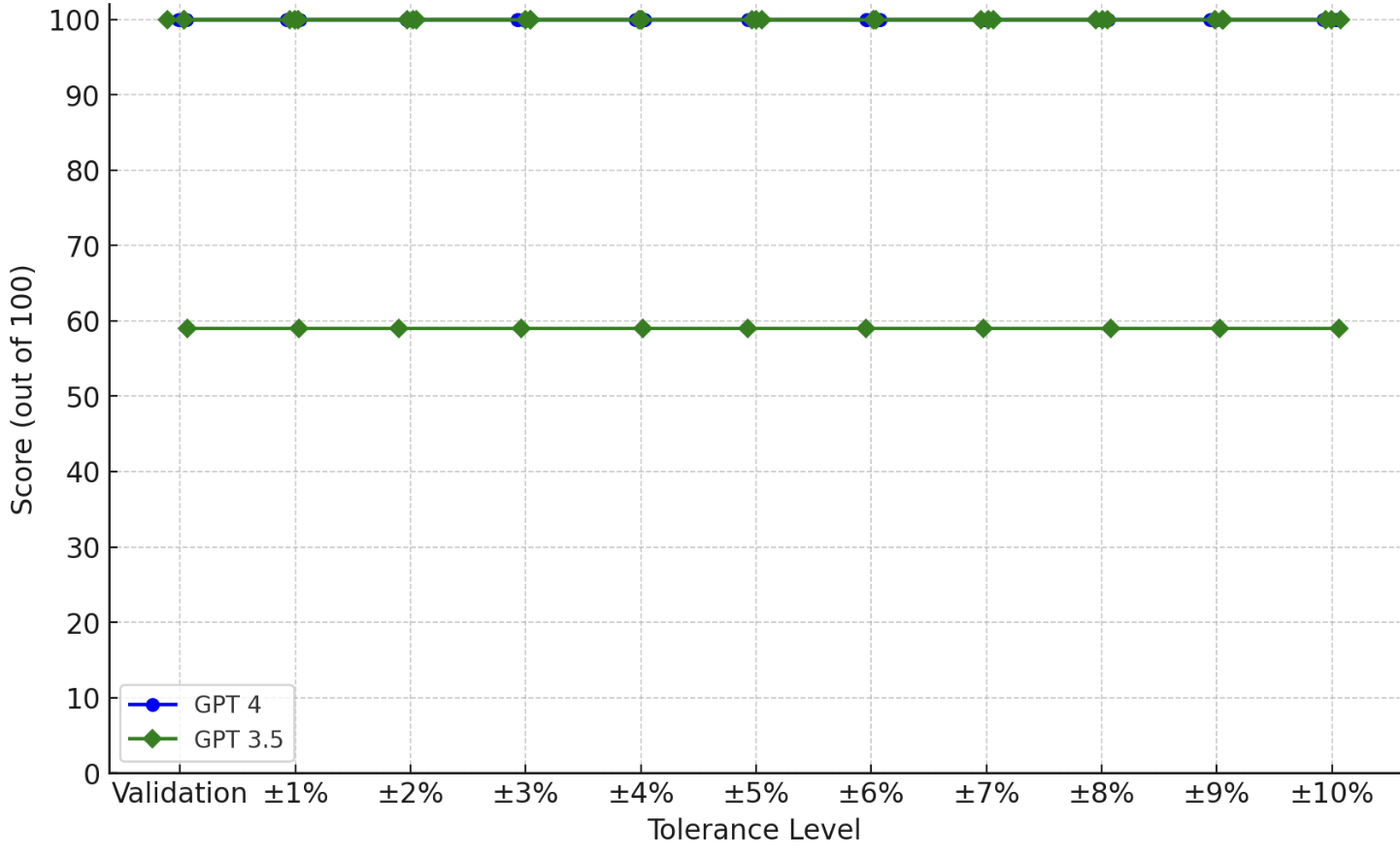}
        \caption{Brackets}
    \end{subfigure}
    \hfill
    \begin{subfigure}{0.32\textwidth}
        \includegraphics[width=\linewidth]{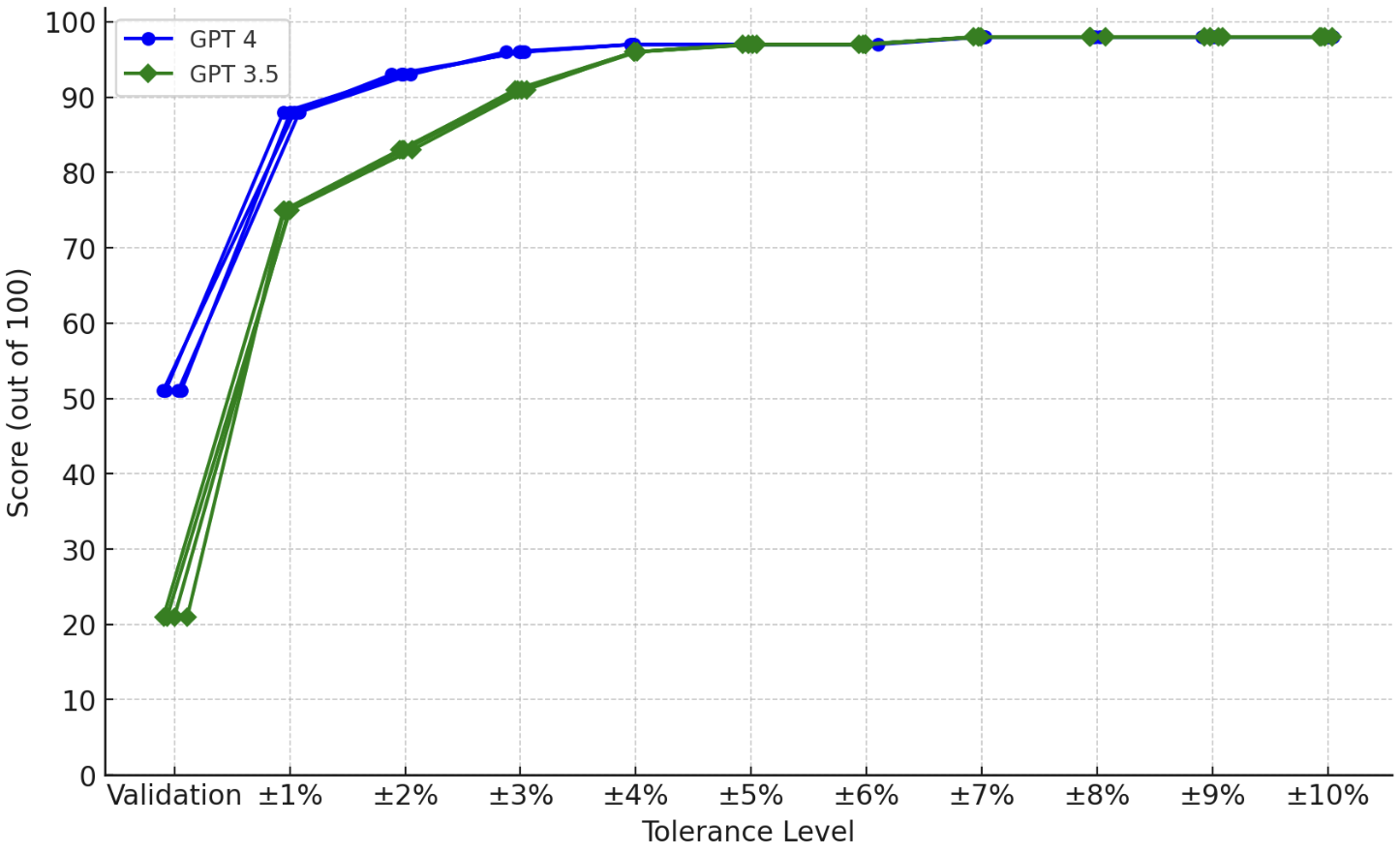}
        \caption{Brackets+Deductions}
    \end{subfigure}
    \hfill
    \begin{subfigure}{0.32\textwidth}
        \includegraphics[width=\linewidth]{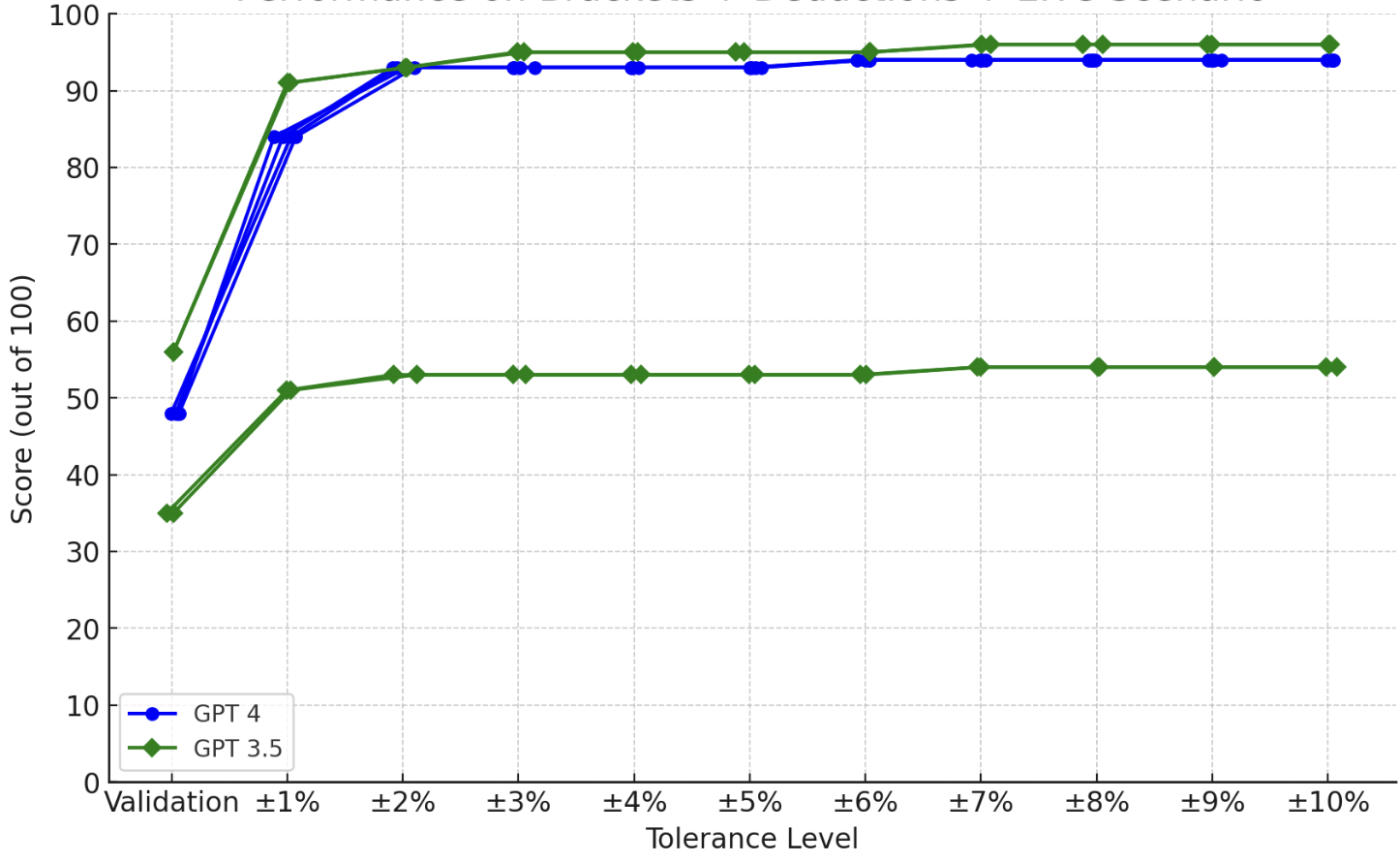}
        \caption{Brackets+Deductions+EITC}
    \end{subfigure}
    \caption{Scenarios with prior code contexts for 4 top ranked candidates per ChatGPT-3.5/4.0.}
\label{fig:accuracy-vs-tolerance-2}
\end{figure}

\vspace{0.25 em}
\noindent \textbf{Results and Discussion}
Table~\ref{table2} shows the results for the top-ranked code candidates generated by ChatGPT-3.5 and ChatGPT-4.0 when provided with prior code context. Comparing Table~\ref{table2} and Table~\ref{table1}, it is evident that providing prior code as a base significantly enhances the performance of LLMs in updating tax software.

\begin{itemize}
    \item \textbf{Overall Strong Performance:} Both GPT-4 and GPT-3.5 exhibit good performance across all scenarios when provided with the previous year's code.  The CodeBertScores are generally high, and MajorityVoteScore is often near-perfect, particularly for GPT-4. This suggests that LLMs can effectively leverage existing code structure to incorporate new tax policy updates.
    
    \item \textbf{GPT-4's Consistent Excellence:}  GPT-4 consistently achieves higher CodeBertScores and accuracy compared to GPT-3.5. In many cases, GPT-4 generates code that achieves both perfect MajorityVoteScore and a perfect match with the ground truth outcomes.
    
    \item \textbf{GPT-3.5's Stroke of Genius:}  Surprisingly, GPT 3.5 showed great performance for the most complicated scenario given prior year's code. We can see it has better top ranked codes than GPT-4. Upon further investigation by looking at charts in Figure~\ref{fig:accuracy-vs-tolerance-2}, we can see that although GPT 3.5 has generated better performing top ranks, but it lacks consistency as it also generated codes that have wrong logic as opposed to GPT-4.0 where if you only look at the Ground Truth Score, it performs worse than GPT3.5 but by looking at charts we can see that the logic of the generated codes via GPT-4.0 are more sound and robust.
    
    \item \textbf{Brackets Only - Near-Perfect Results:}  In the simplest "Brackets Only" scenario, both LLMs excel, with GPT-4.0 consistently achieving perfect results. This indicates that LLMs can easily adapt existing code to update tax brackets with high precision. 
    
    \item \textbf{Lower Ground Truth Matching:}  As scenarios become more complex, the ground truth matching scores decrease, even when MajorityVoteScore remains high. This reveals the presence of subtle errors that might not affect the most frequent output but still deviate from the ideal tax calculation. This suggests that although LLM might be more confident in its generation but that doesn't mean it will generate a code that produces the exact tax in each scenario, emphasizing the need for comprehensive testing methods to detect such nuanced errors. 
    
    \item \textbf{EITC Complexity:} The "Brackets + Deductions + EITC" scenario presents the most significant challenge. While GPT-4.0 maintains high MajorityVoteScore, the ground truth score drops, indicating that EITC logic is still difficult for LLMs to implement accurately, even with prior code context. This suggests that complex tax calculations might require more sophisticated prompting strategies or the integration of additional knowledge sources to guide the LLMs effectively.
\end{itemize}

Table \ref{table2} provides a snapshot of the absolute accuracy and ranking scores of the top code candidates.  However, to assess the robustness of the generated code and its potential for refinement, we analyze the accuracy across a range of error tolerance thresholds. 
The scatter plots in Figure~\ref{fig:accuracy-vs-tolerance-2} visualize the percentage of matching outputs for various tolerance levels when the LLMs are provided with prior year code. Once again, we observe a striking difference in the consistency of GPT-4.0 compared to GPT-3.5.  The generated code by GPT-4.0 consistently achieves near-perfect or perfect accuracy even at stringent tolerance levels. In fact, as highlighted in the introduction, GPT-4.0 achieves 100\% accuracy when allowing an error margin ($\delta$) of at least 7\%, demonstrating its ability to produce code that aligns closely with the ground truth calculations. 

However, a closer look at the scatter plots reveals a nuanced trend: while GPT-4.0 excels at stricter tolerances, GPT-3.5 often exhibits better performance for some generations as the margin of error increases.  For instance, at a tolerance level of 5\% or higher, GPT-3.5 consistently achieves the same or even better accuracy compared to GPT-4.0. Although it may not produce good quality code as often as GPT-4.0, this suggests that GPT-3.5 can leverage the provided code context to generate code that is more robust to larger error margins. 
This observation has important implications for our framework. GPT-3.5, when guided by prior code, might be particularly well-suited for scenarios where a higher tolerance for error is acceptable. Its ability to consistently generate code within a broader acceptable range could be valuable in specific applications. Conversely, GPT-4.0 remains the preferred choice when precision is paramount, as it consistently produces outputs that closely match the ground truth, even at stringent tolerance levels.

These results further validate the effectiveness of our ranking approach. Even in scenarios where the generated code is not perfectly accurate, the ranking system successfully identifies candidates, especially those generated by GPT-4.0, that exhibit high potential for being refined into fully correct implementations.  The scatter plots, by visualizing the accuracy across different error margins, provide insights into the robustness of the generated code and the need for the validation and feedback prompts to achieve the desired level of accuracy.

\section{Discussion}
Since the completion of our initial study that conducted in Fall'23 and Spring'24, focused on ranking code (primarily in \texttt{C} programming language), we have continued to refine our approach to automating tax preparation software updates using Large Language Models (LLMs). A significant advancement is the improvement in model capabilities; even smaller LLMs are now able to accurately modify existing \texttt{Python} code by incorporating new values and adapting to recent tax policy updates. Although this progress is not related to ranking code, it underscores the potential of these models not only in replicating prior implementations but also in generating reliable updates with minimal human intervention. 
In light of these advancements, we are developing a more robust framework aimed at improving the reliability of software updates for tax calculations. This new framework is being tested via symbolic executions across a wider range of LLMs and more complex tax scenarios to assess their ability to autonomously manage code modifications and additions. Preliminary results indicate that even smaller LLMs can achieve high accuracy in updating and extending \texttt{Python} code, which is promising for the future of automated tax software maintenance. The ongoing work is expected to significantly aid tax prep software developers to update their code as the tax law evolves every year.

\section{Conclusion}
\label{sec:conclusion}
The ever-growing complexity of tax law and policies has significantly increased the role of tax preparation software in navigating the intricacies of legal accountability and compliance. 
As the tax law gets updated, maintaining the compliance and trustworthiness of tax prep software is challenging. 
As part of a wider NSF-sponsored project, our goal is to develop principled techniques and tools to support software programmers in maintaining tax preparation software. 
Our framework combines best practices from formal methods (metamorphic specifications), software engineering (automated testing and debugging), and AI (LLMs for code-generation) to ensure that the software not only adheres to the latest tax regulations but also remains easy-to-maintain.
By leveraging this integrated approach, we aim to reduce the time and effort required for updates, enhance the accuracy of tax calculations, and ultimately improve the reliability and user trust in tax preparation software. 
This will enable programmers to more effectively respond to legislative changes and meet the needs of taxpayers efficiently.

\section*{Acknowledgements}
This project has been partially supported by the NSF under grants CCF-$2317206$ and CCF-$2317207$.

\bibliographystyle{abbrv}
\bibliography{paper}

\end{document}